\documentclass[12pt,preprint]{aastex}

\slugcomment{{\it The Astrophysical Journal}}

\shorttitle{A hard X-ray view of two distant VHE blazars}
\shortauthors{Reimer et al.}

\begin{document}

\title{A hard X-ray view of two distant VHE blazars: 1ES 1101-232 and 1ES 1553+113}

\author{A. Reimer\altaffilmark{1}, L. Costamante\altaffilmark{1}, G. Madejski\altaffilmark{2}, O. Reimer\altaffilmark{1} \and D. Dorner\altaffilmark{3}}

\altaffiltext{1}{W.W. Hansen Experimental Physics Laboratory \& Kavli Institute for Particle Astrophysics \& Cosmology, Stanford University, 452 Lomita Mall, Stanford, CA 94305, USA; afr@stanford.edu, Luigi.Costamante@stanford.edu, olr@stanford.edu}

\altaffiltext{2}{Kavli Institute for Particle Astrophysics and Cosmology, Stanford University, Stanford, CA 94305, and Stanford Linear Accelerator Center, 2575 Sand Hill Road, Menlo Park, CA 94025, USA; madejski@slac.stanford.edu}

\altaffiltext{3}{Universit\"at W\"urzburg, D-97074 W\"urzburg, Germany; dorner@astro.uni-wuerzburg.de}

\begin{abstract}
TeV-blazars are known as prominent non-thermal emitters across the 
entire electromagnetic spectrum with their photon power peaking in the X-ray 
and TeV-band. If distant, absorption of $\gamma$-ray photons by the 
extragalactic background light (EBL) alters the intrinsic TeV spectral 
shape, thereby affecting the overall interpretation.
Suzaku observations for two of the more distant
TeV-blazars known to date, 1ES~1101-232 and 1ES~1553+113, 
were carried out in May and July 2006, respectively,
including a quasi-simultaneous coverage with the 
state of the art Cherenkov telescope facilities.
We report on the resulting data sets with emphasis on the X-ray band, and set
into context to their historical behaviour.
During our campaign, we did not detect any significant X-ray or $\gamma$--ray variability.  
1ES~1101-232 was found in a quiescent state with the lowest
X-ray flux ever measured. The combined XIS and HXD PIN data 
for 1ES~1101-232 and 1ES~1553+113 clearly indicate spectral curvature up to 
the highest hard X-ray data point ($\sim 30$ keV), 
manifesting as softening with increasing energy.
We describe this spectral shape by either a broken power law or a log-parabolic fit with
equal statistical goodness of fits.
The combined 1ES 1553+113 very high energy spectrum ($90-500$~GeV) did not show any significant changes with respect
to earlier observations. 
The resulting contemporaneous broadband spectral energy distributions of both TeV-blazars
are discussed in view of implications for intrinsic blazar parameter values, taking into
account the $\gamma$-ray absorption in the EBL.

\end{abstract}

\keywords{X-rays: galaxies - galaxies: active - BL Lacertae objects: individual (1ES 1101-232, 1ES 1553+113) }

\section{Introduction}

Blazars are among the most extreme sources in the high energy sky. They constitute a subclass
of the jet-dominated radio-loud population of active galactic nuclei (AGN), which in turn are sub-divided into
flat spectrum radio quasars (FSRQ) and BL Lac objects where the absence or depression of strong emission lines
characterizes the latter. Their observational properties include
non-thermal continuum emission and irregular variability across the whole electromagnetic band, ranging from 
subhours at gamma-ray energies to week-long time scales in the radio band, 
often high optical and radio polarization,
and a core-dominated radio morphology. In some sources, superluminal motion has been detected, indicating
a relativistically enhanced emission region close to the line-of-sight \cite{Blandford78}.
As a consequence of this, any jet emission is highly beamed.

The spectral energy distribution (SED) of blazars 
shows two broad peaks in the $\nu F_\nu$ representation. The low energy hump is generally agreed to stem
from synchrotron radiation of relativistic electrons and/or positrons in the jet whereas the origin of 
the high energy one is still
under debate.  Leptonic models explain the complete non-thermal SED as
synchrotron and inverse Compton emission from
relativistic electrons or pairs which upscatter their self-produced
synchrotron radiation or external
photon fields. Hadronic models produce the high 
energy peak either via interactions of relativistic protons with matter
\cite{Pohl00}, 
ambient photons \cite[e.g.][]{Mannheim93}, magnetic fields \cite{Aharonian00}, 
and/or both, magnetic fields and photons \cite[e.g.][]{Muecke01,Muecke03}.

Almost all of the 19 blazars (except for BL Lacertae \cite{Albert07c} and 3C~279 \cite{Teshima07}) detected 
reliably to date at the TeV energy belong to the subclass
of high-frequency peaked BL Lacs (HBLs), with their two $\nu F_\nu$ peaks typically
at UV/X-rays and TeV-energies. Until recently blazars have been detected preferentially in their
flare state, owing to instrumental limitations. With the contemporary 
Cherenkov telescopes (such as H.E.S.S., MAGIC, VERITAS) it is possible to study
also low activity states (subsequently referred to "quiet state").

TeV-blazars have also been used successfully as a powerful tool to probe the extragalactic
background light (EBL) at IR/optical energies via the absorption of 
$\gamma$-rays in the cosmic diffuse radiation field. 
The most stringent constraints on the EBL density
at $\sim 2\mu$m stem from observations with the H.E.S.S. 
telescope system of the high-redshift, hard-spectrum TeV-blazar 
1ES~1101-232 \cite{Aharonian06a}. The most distant (redshift $z>0.1$) TeV-blazars 
established to date are 1ES~1101-232 ($z=0.186$),
1ES~1218+304 ($z=0.182$), 1ES~1011+496 ($z=0.212$), 1ES~0347-121 ($z=0.185$), 
1ES~0229+200 ($z=0.140$), H~2356-309 ($z=0.165$), 
H~1426+428 ($z=0.129$), PKS~2155-304 ($z=0.116$), 1ES 1553+113 
\cite[$z>0.25$:][]{Perlman96,Falomo90,Aharonian06b}.

Extreme HBLs, as is the case for most very high energy (VHE) blazars, 
are characterized by their spectral extension to extremely high particle energies.
TeV-blazars have been regularly observed in the past by X-ray instruments, however, 
not many have been detected so far in 
the hard X-ray range where the highest energy particles leave their imprints. 
The sensitivity reached by the hard X-ray detector HXD onboard the Suzaku 
satellite allowed for the first time to study in detail possible new spectral 
features and curvature at $>10$~keV, even in a blazar quiet state, which may 
help understanding the properties and origin of the VHE emission.

We report here on our $\sim 50$~ksec Suzaku observations on each 
of two more distant TeV-blazars known to date: 1ES~1101-232 and 1ES~1553+113.
These observations were covered by quasi-simultaneous observations with 
the H.E.S.S. telescope system and the MAGIC telescope.
None of these sources have shown in the past significant variability in the TeV-band
\cite{Aharonian06a,Aharonian06b},
likely owing to their low flux close to the detection limit of the TeV facilities.

The paper is organized as follows: We provide a brief overview of both TeV-blazars 
in Sect.~2. Sect.~3 is devoted to
a technical description of our Suzaku data analysis and its results. In Sect.~4
we report on the quasi-simultanuous multiwavelength coverage during the Suzaku observations of
1ES 1101-232 and 1ES 1553+113. In Sect.~5 we discuss our results in particular in context
to their historical behaviour. Sect.~6 summarizes the main results of this work.

%%%%%%%%%%%%%%%%%%%%%%%%%%%%%%%%%%%%%%%%%%%%%%%%%%%%%%%%%%%%%%%%%%%%%%%%%%%%%%%%%%%%%%%%%

\section{The VHE blazars 1ES~1101-232 and 1ES~1553+113}

The X-ray selected BL Lac object (XBL) 1ES~1101-232, hosted by an 
elliptical galaxy at redshift $z=0.186$ with estimated magnitude $m_R=16.41$, one 
of the brightest BL Lac hosts known so far, is presumably located in a galaxy cluster.
It has been detected by various X-ray instruments: the Ariel-5 X-ray satellite 
discovered this source, HEAO-1, Einstein, ROSAT, BeppoSAX, RXTE, XMM, SWIFT, 
Suzaku (this paper) just to name a few. Its strong X-ray-to-optical flux 
ratio led to the classification of an XBL \cite{Scarpa97}, which has later 
been revised to a high-frequency peaked BL Lac object \cite{Donato01} within the HBL-LBL 
scheme \cite{Giommi95}.  The radio maps of 1ES 1101-232 show a one-sided, 
not well-collimated jet structure at a few kpc distance from the core 
\cite{Laurent-Muehleisen93}. The optical core emission is typically 
varying on time scales of months \cite[e.g.][]{Remillard89}, although
occasional intraday flares have been reported \cite{Romero99}. 
Based on the typical broadband characteristics of VHE AGN several 
authors have predicted this source to be a TeV-blazar 
\cite[e.g.][]{Wolter00,Costamante02}. Indeed, it was recently 
detected by the H.E.S.S. Cherenkov telescope system \cite{Aharonian06b}.
At MeV-GeV energies, however, EGRET reported only an upper limit from 1ES~1101-232 \cite{Lin96}.

Owing to the non-detection of any lines nor resolving its host galaxy 
the redshift of the XBL/HBL 1ES 1553+113 is to date still unknown.
Redshift limits, though, have been imposed either through HST 
results \cite[$z>0.78$:][]{Sbarufatti05} on the assumption that the nucleus' host
is a typical BL Lac host, or through arguments involving TeV photon 
absorption in the EBL (\cite[$z>0.25$, $z<0.74$:][]{Aharonian06b}; 
\cite[$z<0.42$:][]{Mazin07}) and no anomalies in the source intrinsic 
high energy SED. Like 1ES 1101-232, 1ES 1553+113 is a strong X-ray source 
and has been visited by several X-ray instruments, however, focusing mainly
towards the soft part of its X-ray spectrum (Einstein, ROSAT, BeppoSAX, 
RXTE, XMM). Its weak radio flux \cite[a few tens of a mJy at 5GHz:][]{Falomo90} 
among the blazar class leads to its classification of an XBL and HBL \cite{Falomo90,Donato05}. 
Results from a 3-weeks multifrequency campaign in 2003 covering the 
radio-to-X-ray range \cite{Osterman06} where a strong X-ray flare was detected,
indicate an extremely high synchrotron cutoff energy, superceding in this 
respect even well-known extreme HBLs like Mkn 421, Mkn 501 
or PKS 2155-304. This places 1ES~1553+113 right at the frontline of 
the class of extreme HBLs \cite{Rector03}. Despite its unknown redshift, it was 
considered as a promising source for detection at TeV energies based 
on its spectral broad band properties.  The H.E.S.S. collaboration 
reported recently its discovery at VHEs with a 
integral flux detection on a $3.8\sigma$ level, and 
a very soft spectrum \cite{Aharonian06b}. This
was confirmed by MAGIC \cite{Albert07b}.
The Third EGRET catalog \cite{3EG} does not list this blazar 
as identified in the MeV-GeV energy band.

\section{Suzaku observations and data analysis}

The two objects have been observed by Suzaku's X-ray Imaging Spectrometer \cite[XIS:][]{Koyama07}
and the Hard X-ray Detector \cite[HXD:][]{Takahashi07} in May 25-27 2006 
for a total of 53.1~sec on-source for 1ES 1101-232, and in July 24-25 2006 
for a total of 41.1~sec on-source for 1ES 1553+113. XIS is a CCD instrument with
three front-illuminated and one back-illuminated CCD camera at the focal plane of the four X-ray telescopes 
\cite[XRT:][]{Serlemitsos07}, and sensitive in the energy range 0.2-12 keV. 
During our observations of the two TeV-blazars, XIS was operated in the 
$3\times 3$ and $5\times 5$ observation modes. No charge injection 
procedures were carried out during these observations. 

The HXD consists of an array of $4\times 4$ detectors (well units) and 20 surrounding crystal scintillators for 
active shielding. Each unit consists of four Si PIN diodes (PIN), sensitive in the 10-70 keV energy range, and 
four GSO/BGO phoswitch counters (GSO) for detecting photons in the 40 - 600 keV regime. 
Since the focus of these observations were the hard part of the X-ray energy regime, the HXD was chosen as the 
nominal point to ensure optimal sensitivity.

One day before our Suzaku observations of 1ES 1101-232 started, the HXD team started changing the bias voltage, 
affecting potentially HXD's UNIDIDs 0 to 3, to suppress the rapid 
increase of noise events noted at this time.
In particular, on May 26 2006, the bias voltage of the HV-P0 was increased in two steps, affecting
the measurements of the PIN during the 1ES 1101-232 observations: we found two huge background 'flares' in 
UNIDID 1, towards the end of this observation beyond MJD~53941.7.

Our data analysis is based on the version 2 processed data with the standard mkf filtering, and using
tools of the HEASoft version 6.4. The CALDB release from February 1 2008 contains the calibration files 
used in this analysis.
For the data reduction we followed the ``The Suzaku Data Reduction Guide''\footnote{http://suzaku.gsfc.nasa.gov/docs/suzaku/analysis/abc/} 
and specifically Ver. 2.0 if not noted otherwise.

%%%%%%%%%%%%%%%%%%%%%%%%%%%%%%%%%%%%%%%%%%%%%%%%%%%%%%%%%%%%%%%%%%%%%%%%%%%%%%%%%%%%%%%%%%%%%%%%%%%%%%%%%%%%%%%%

\subsection{XIS data reduction and analysis}

We started the XIS data analysis with the unscreened version 2 of the 
data.  We computed the XIS pulse invariant and grade values and updated the 
CTI calibration via the tool {\tt xispi}.  The 
subsequent reduction, including cleaning of hot and 
flickering pixels, selection of grades 0, 2, 3, 4, and 6, filtering out the 
epochs of high background (e.g. due to the satellite crossing through 
the South Atlantic Anomaly) was performed using the tool {\tt xselect} 
(and specifically a script {\tt xisrepro}.  
Source events were then extracted in a circle around the source 
with radius of $\sim 260\arcsec$. The extraction of background 
events were performed from a region devoid of any obvious X-ray sources.
We ensured that the use of other background regions within the 
XIS field does not significantly affect
the results from our spectral analysis. 
We generated the corresponding response (rmf) and auxillary (arf) files by utilizing the 
tools {\tt xisresp} and {\tt xissimarfgen}, using 
the default energy steps;  the latter is a 
Monte-Carlo code provided by the Suzaku team, and it also accounts for the 
hydrocarbon contamination that has been built-up in the optical 
path of each XIS sensor. 
With the most recent calibration the spectra do not show calibration anomalies down to $\sim 0.4$~keV, which we therefore included into the analysis. Our spectral results do not change significantly 
even when limiting the XIS bandpass to above 0.7~keV.

%%%%%%%%%%%%%%%%%%%%%%%%%%%%%%%%%%%%%%%%%%%%%%%%%%%%%%%%%%%%%%%%%%%%%%%%%%%%%%%%%%%%%%%%%%%%%%%%%%%%%%%%%%%%%%%

\subsection{HXD data reduction and analysis}

As was the case for the XIS, here we also used the 
version 2 processed data.  Here, we started with the filtered data, where
standard cleaning criteria were used;  however, for the extraction of the 
spectrum of 1ES 1101-232 we selected only the data before MJD~53941.7 to avoid
any contamination with the two huge background flares.  We also extracted 
the non-X-ray background spectrum from the background files provided by HEASARC, 
but applied the same time selection criteria as to the source files.  
The source spectrum was dead-time corrected using the tool {\tt hxddtcorr};  we note that the 
application of {\tt hxddtcorr} resulted in a live time of 93.4\% for the 1ES 1101-232 and 
93.8\% for the 1ES 1553+113 observations.  
We applied analogous procedure to the GSO data, but since neither source was detected in the GSO, we did
not use those data in the subsequent analysis.  

For the spectral analysis of the combined time-integrated XIS and PIN data
we have taken into account the cosmic X-ray background \cite{Gruber99}
normalized to the PIN field-of-view,
and a XIS/PIN normalization of 1.12 \cite{Ishida07}.

%%%%%%%%%%%%%%%%%%%%%%%%%%%%%%%%%%%%%%%%%%%%%%%%%%%%%%%%%%%%%%%%%%%%%%%%%%%%%%%%%%%%%%%%%%%%%%%%%%%%%%%%%%%%%%

\subsection{Results}

We detected 1ES~1101-232 and 1ES~1553+113 with Suzaku up to energies above $\sim$30~keV.
This is the highest X-ray energy measurement so far obtained for 1ES~1553+113.

\begin{figure}[t]
{\rotatebox{-90}{\includegraphics[height=18.cm]{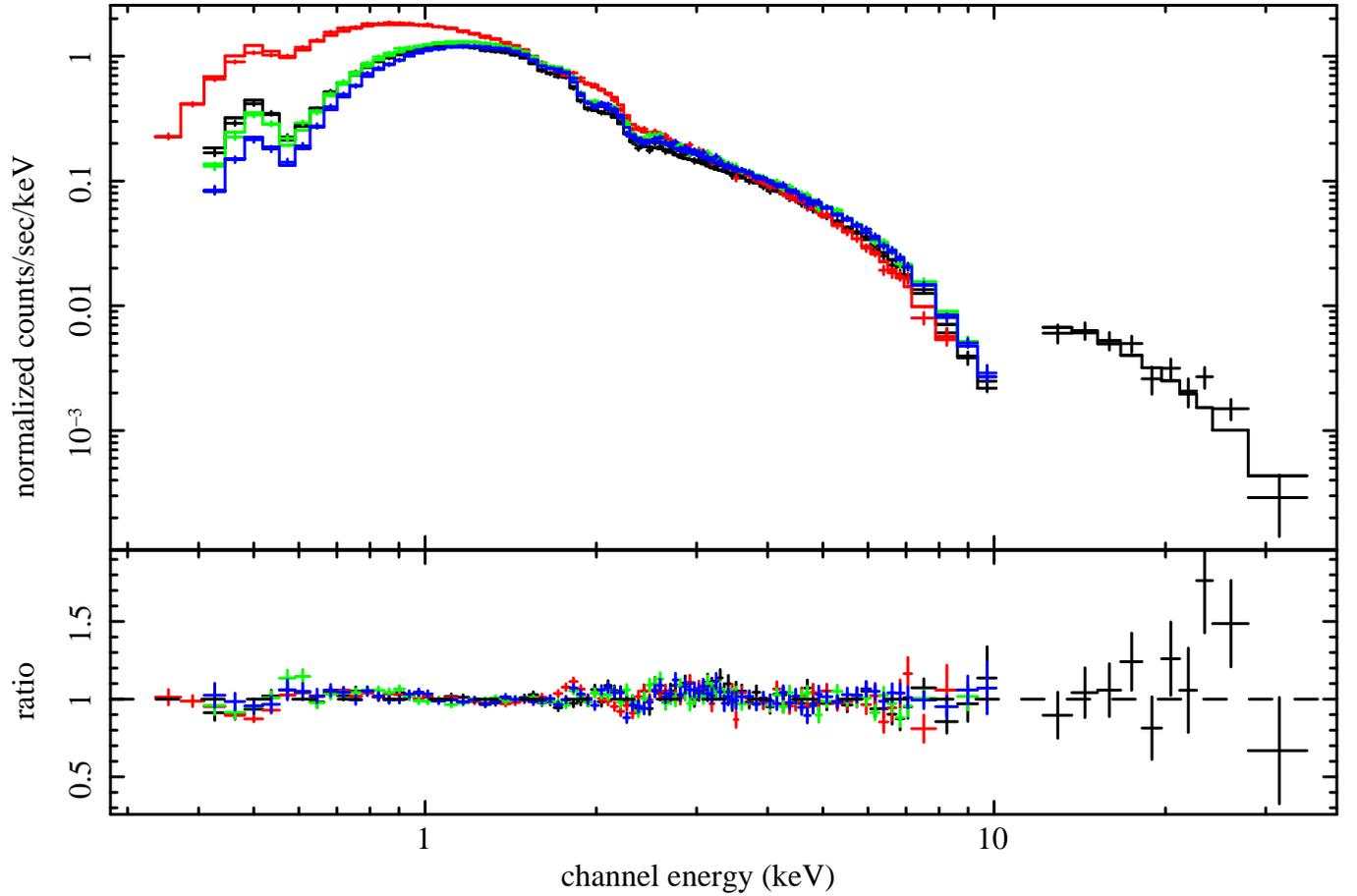}}}
\caption{Data and folded broken power law model of 1ES~1101-232 that fits the data best ($\chi_{\rm red}^2=1.11$) and residuals.
Parameters:
$\Gamma_1=2.04\pm 0.02$, $\Gamma_2=2.32\pm 0.02$, $E_{\rm break}=1.37\pm 0.08$keV. 
The four curves below 10~keV correspond to the four XIS detectors while the curve above 10~keV denote the PIN data.}
\label{fig1}
\end{figure}

The combined XIS-PIN data has been fitted to both, simple and broken power laws.
For both sources we found a significantly improved $\chi^2$-value when fitting to 
a broken power law as compared to a single power law. 
The absorbing column densities were fixed to the respective Galactic values 
here ($N_H=5.76\cdot 10^{20}$~cm$^{-2}$ for 1ES~1101-232, $N_H=3.67\cdot 10^{20}$~cm$^{-2}$ 
for 1ES~1553+113).  The best-fit parameters and goodness-of-fits are summarized 
in Table~\ref{tab1}.  Both sources show the break to be around 1.4~keV with 
1ES~1553+113 having a somewhat steeper overall spectrum than 1ES~1101-232.
We determined the average 2-10 keV flux from 1ES~1101-232 to be $1.68\cdot 10^{-11}$~erg 
cm$^{-2}$s$^{-1}$, and the 10-30 keV flux to be $7.4\cdot 10^{-12}$~erg cm$^{-2}$s$^{-1}$ 
(source flux only).  This is the lowest flux level so far measured from this source.
The average 2-10 keV source flux from 1ES~1553+113 is $3.54\cdot 10^{-11}$~erg 
cm$^{-2}$s$^{-1}$, about a factor 2 higher than for 1ES~1101-232. The
10-30 keV flux is at $1.35\cdot 10^{-11}$~erg cm$^{-2}$s$^{-1}$.
This X-ray flux level is comparable to the 2001 XMM data, and 
lies at an intermediate value between the so far highest (found 
during the 2005 SWIFT observations) and lowest (detected during 
the 2003 RXTE campaign) flux state of 1ES~1553+113. 
The best-fit and residuals for both sources are shown in Fig.~\ref{fig1} and \ref{fig4}.

\begin{figure}[t]
\resizebox{\hsize}{!}{\includegraphics{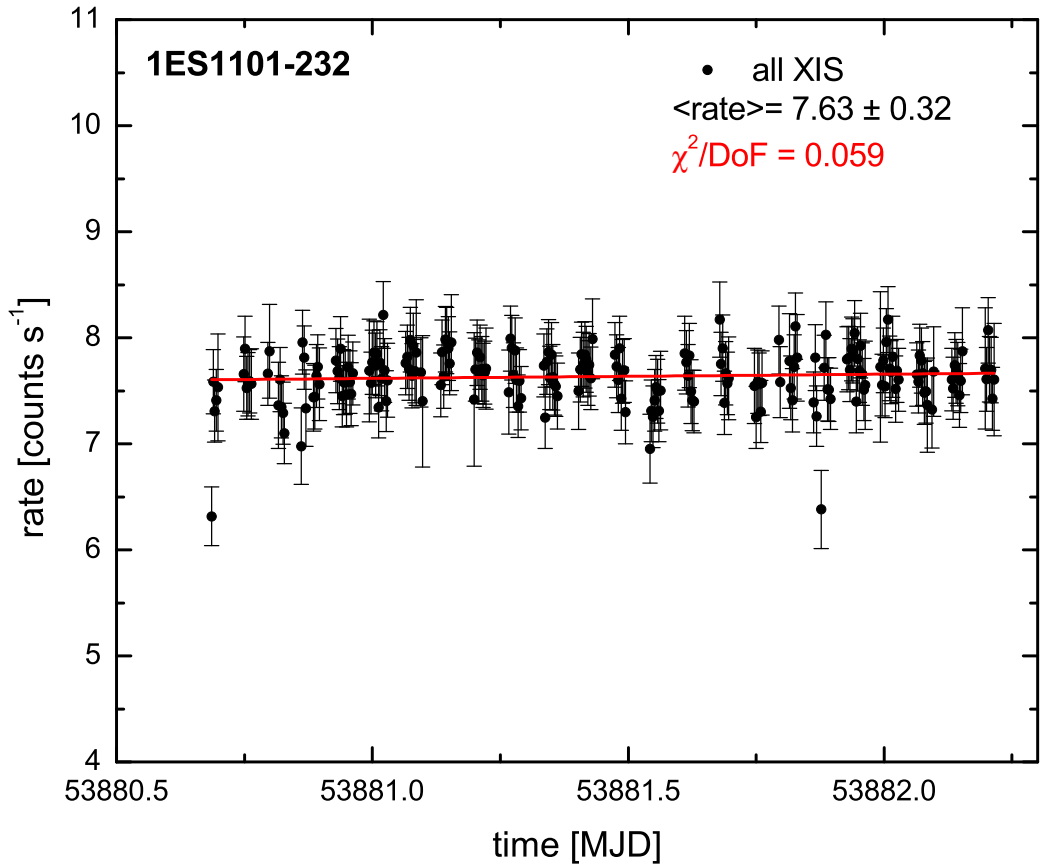}}
\caption{Background corrected lightcurve of XIS data sampled with 256~sec binning. A $\chi^2$ fit to
a straight line indicates no signatures of variability of the source during the Suzaku observations.}
\label{fig2}
\end{figure}

\begin{figure}[t]
%\rotatebox{-90}{{\includegraphics[height=18.cm]{f3.eps}}}
\resizebox{\hsize}{!}{\includegraphics{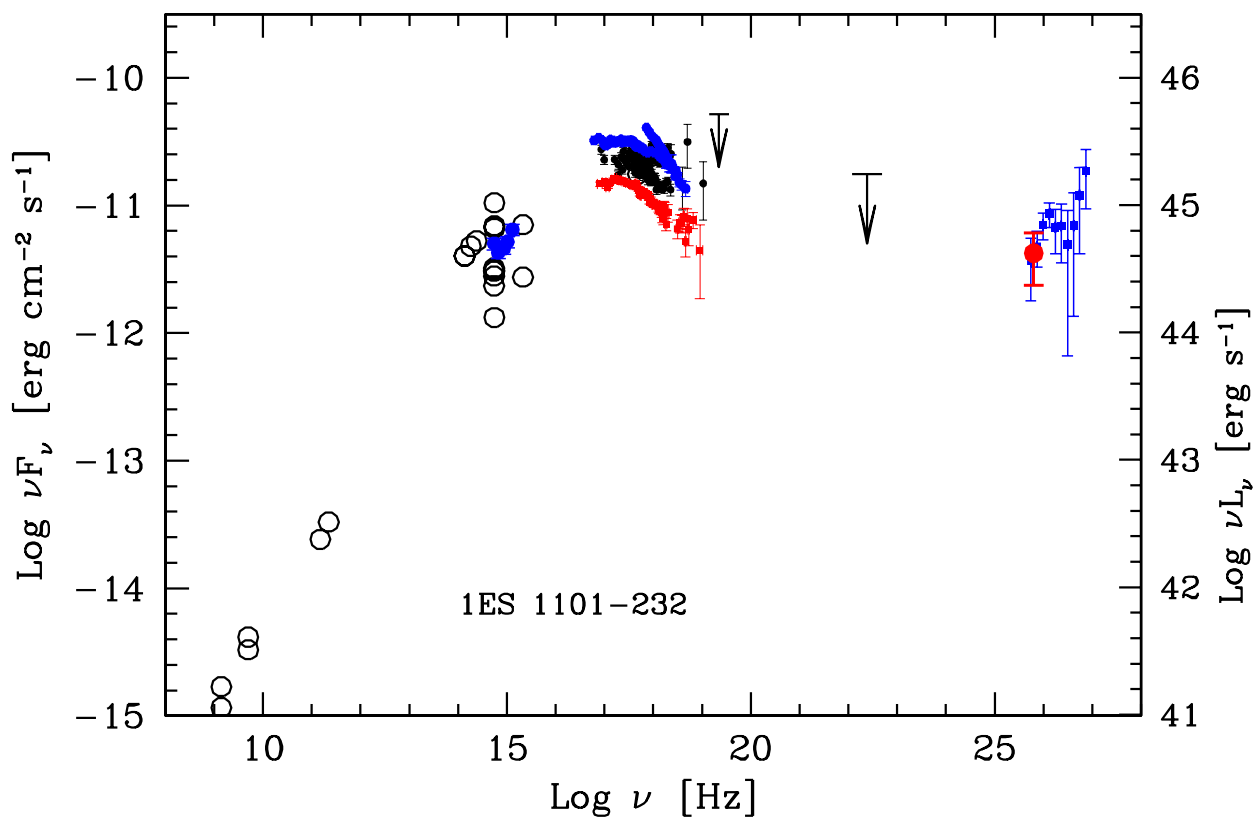}}
\caption{Data and folded best fit ($\chi_{\rm red}^2=1.08$) broken power law model of 1ES~1553+113.
Parameters: $\Gamma_1=2.13\pm 0.02$, $\Gamma_2=2.42\pm 0.01$, $E_{\rm break}=1.38\pm 0.06$~keV.
Note that the PIN data lie systematically below the model indicating a further steepening. For a description of the individual curves see Fig.~\ref{fig1}.}
\label{fig4}
\end{figure}

\begin{figure}[t]
%\resizebox{\hsize}{!}{\includegraphics{f4.eps}}
\rotatebox{-90}{{\includegraphics[height=18.cm]{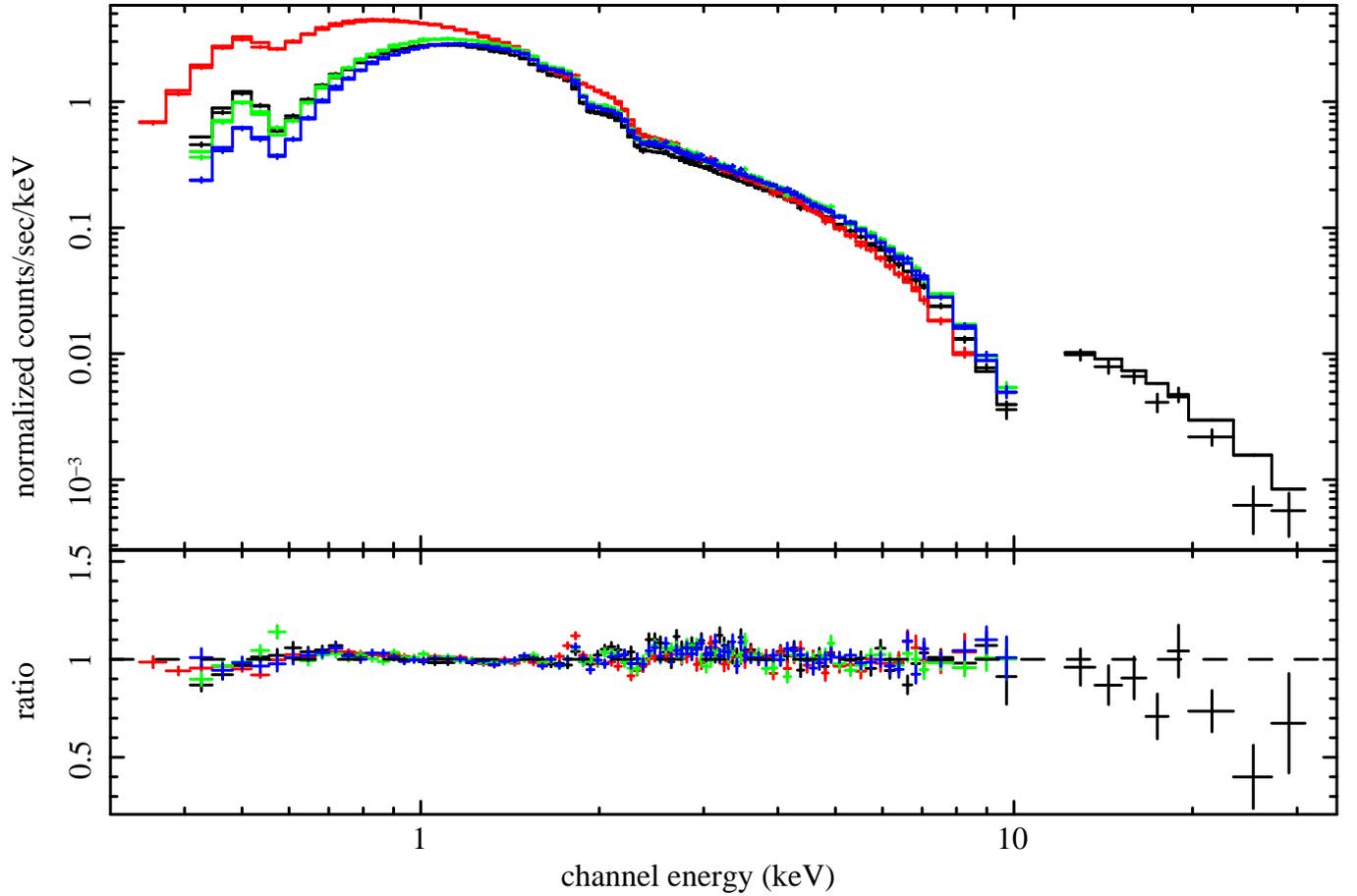}}}
\caption{Background corrected XIS lightcurve sampled with 256~sec binning. 
A $\chi2$-test indicates a slight preference of representing the XIS lightcurve 
with a linear fit better than with a constant flux.
We found no signatures of short time variability during the Suzaku observations.} 
\label{fig5}
\end{figure}

We tested for variability in both Suzaku data sets using the $\chi^2$ method.
No flux variability on subhour or intraday time scales were found (Fig.~\ref{fig2}, 
\ref{fig5}). The lack of variability, otherwise often observed on
these time scales during high flux states in TeV-blazars such as 
Mkn~501, Mkn~421, PKS 2155-304, etc., may suggest 
the sources to have lingered at a quiescent activity level in 2006.

Log-parabolic shapes have been proposed to provide a better description of 
X-ray spectral shapes in HBL-type AGN \cite[e.g.][]{Perlman05}. A $\chi^2$-test to the shape 
$dN/dE \propto E^{[-\Gamma-\beta\log(E)]}$ of the combined XIS-PIN 
spectrum revealed a statistically comparable goodness of fit as compared 
to the broken power law representation up to the highest reported energy point 
for both sources. Best-fit parameters for the log-parabolic shapes are in Table~\ref{tab1}.
Although the best-fit curvature value for both sources is similar, 
we note a tendency of a further steepening in the PIN data with respect to the model
(log-parabolic or broken power law shape)
describing the 1ES~1553+113 data set (see e.g. Fig.~\ref{fig4}).
A log-parabolic shape spanning the whole decline of the 
synchrotron component can be the result of energy-dependent 
acceleration \cite[e.g.][]{Protheroe99,Massaro04}, episodic \cite{Perlman05} 
or fluctuating particle energy gains \cite{Trama07}. If this 
interpretation holds, the present result indicates acceleration operating 
also in a quiet state of source activity. According to Protheroe \& Stanev 
(1999) the spectral shape of 1ES~1101-232 is in agreement with a cutoff 
due to either $E^2$ particle energy losses or a finite size of the emission region for
an energy dependence of acceleration and/or escape $\propto E^{-1/3}$ or weaker.

Our investigation of the Suzaku data found no convincing signatures of spectral 
hardening up to the highest detected X-ray energy point, as would be 
expected from a synchrotron emitting Compton loss dominated source 
in an energy range affected by the Klein-Nishina decline of the cross 
section, or the onset of the $\gamma$-ray component. We therefore conclude 
that the measured X-ray radiation is entirely due to the synchrotron component,
and that losses are either not Compton loss dominated in the X-ray band, 
or are still in the Thomson loss regime. 

\clearpage

\clearpage

\begin{table*} 
\begin{center}
  \begin{tabular}{l|llll|lll} 
\hline
{\small Object}                  & {\small $\Gamma_1$} & {\small $\Gamma_2$} & {\small $E_{\rm break}$} & {\small $\chi_{\rm red}^2$} & {\small $\Gamma$} & {\small $\beta$} & {\small $\chi_{\rm red}^2$} \\
                        &            &            &  {\small $\mathrm{keV}$} &   {\small   (d.o.f.)}      &          &         &  {\small (d.o.f.)}         \\ \hline
{\small 1101-232}            & {\small $2.04\pm 0.02$}& {\small $2.32\pm 0.02$}& {\small $1.37 \pm 0.08$} & {\small 1.11 (3328)}   & {\small $2.10 \pm 0.01$} & {\small $0.26 \pm 0.02$} & {\small 1.10 (3329)}\\         
{\small 1553+113}            & {\small $2.13 \pm 0.02$} & {\small $2.42\pm 0.01$} & {\small $1.38\pm 0.06$} & {\small 1.08 (4354)} & {\small $2.19 \pm 0.01$} & {\small $0.26 \pm 0.01$} & {\small 1.06 (4355)}\\ \hline
  \end{tabular}
\end{center}
\caption{Spectral fit results from the combined XIS-PIN Suzaku data of a broken power law fit with photon indices $\Gamma_1$, $\Gamma_2$ and break 
energy $E_{\rm break}$ and of a log-parabolic shape with photon index $\Gamma$ and curvature $\beta$. Galactic column densities has been used for
these fits.
}
  \label{tab1}
\end{table*}
\clearpage

%%%%%%%%%%%%%%%%%%%%%%%%%%%%%%%%%%%%%%%%%%%%%%%%%%%%%%%%%%%%%%%%%%%%%%%%%%%%%%%%%%%%%%%%%%%%%%%%%%%%
%%%%%%%%%%%%%%%%%%%%%%%%%%%%%%%%%%%%%%%%%%%%%%%%%%%%%%%%%%%%%%%%%%%%%%%%%%%%%%%%%%%%%%%%%%%%%%%%%%%%

\section{Simultaneous multifrequency observations}

%%%%%%%%%%%%%%%%%%%%%%%%%%%%%%%%%%%%%%%%%%%%%%%%%%%%%%%%%%%%%%%%%%%%%%%%%%%%%%%%%%%%%%%%%%%%%%%%%%%%%%%%%%%%%%%%%%%%%
\subsection{Observations in the optical band}
 
The KVA observatory on La Palma is monitoring 1ES 1553+113 regularly since early 2006. KVA observed this source
between 21 and 27 July 2006 in the R-band \cite{Albert07a}. The KVA pointing date that falls into the Suzaku observing time is MJD=53941.43495.
The measured flux was corrected for Galactic extinction using a R-band extinction of $A=0.172$~mag \cite{Schlegel98}.
Since a host galaxy contribution for 1ES~1553+113 is negligible, no host galaxy correction for the flux has been made. A flux value of 
$S_{\rm opt} = 14.28 \pm 0.23$~mJy was derived.

%%%%%%%%%%%%%%%%%%%%%%%%%%%%%%%%%%%%%%%%%%%%%%%%%%%%%%%%%%%%%%%%%%%%%%%%%%%%%%%%%%%%%%%%%%%%%%%%%%%%%%%%%%%%%%%%%%%%%
\subsection{Simultaneous TeV-observations}

1ES 1101-232 was visited by the High Energy Stereoscopic System (H.E.S.S.) atmospheric Cherenkov 
telescope array several times between
2004 and 2007 \cite{Aharonian07a,Aharonian06a,Benbow07}. In May 2006 H.E.S.S. observations were scheduled to 
cover the epoch of the Suzaku measurements towards 1ES 1101-232. The 4.3~h good-quality data revealed
only a marginal detection of $2.9\sigma$ (51 excess events) and a corresponding
flux of $I(>260{\rm GeV}) = 3.2\pm1.4_{\rm stat}\times 10^{-12}$cm$^{-2}$s$^{-1}$ \cite{Benbow07,Aharonian07c}. 
The complete 2006 data set showed a $3.6\sigma$ excess in 13.7~h live time.
In comparison, the 2004-2005 TeV-data on this source revealed an integral flux of 
$I(>200{\rm GeV}) = 4.5\pm1.2\times 10^{-12}$cm$^{-2}$s$^{-1}$ ($10.1\sigma$) \cite{Aharonian07a}.
The low X-ray flux level during the Suzaku compaign was therefore accompanied with a comparable
VHE flux when compared to the 2004/05 activity state. All 2004/05 VHE data are compatible with a
differential power law spectrum $dN/dE = (5.63 \pm 0.89) \cdot 10^{-13} {\rm (E/TeV)}^{-(2.94\pm 0.20)}$cm$^{-2}$ s$^{-1}$ TeV$^{-1}$ 
\cite{Aharonian07a}.

\begin{figure}[t]
\resizebox{\hsize}{!}{\includegraphics{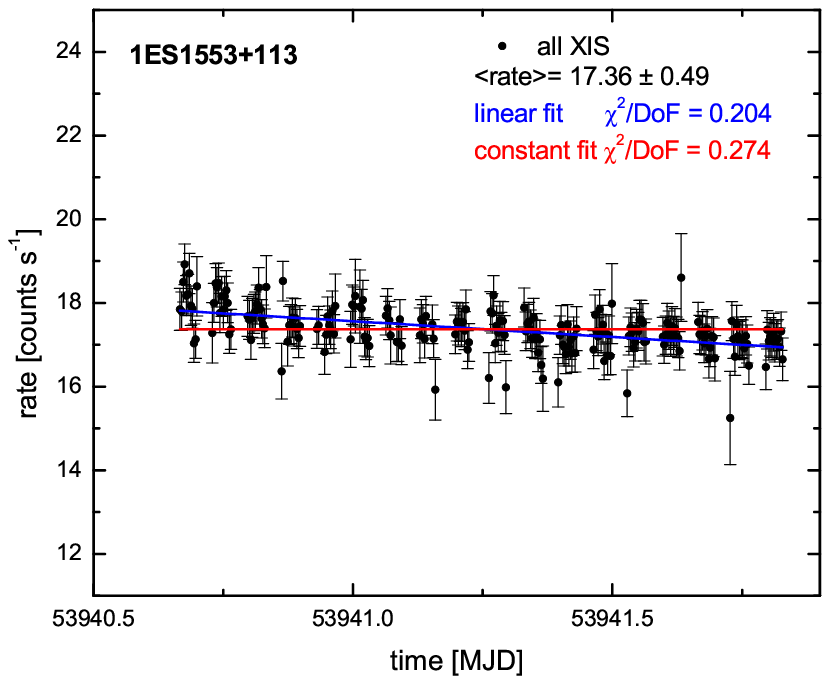}}
\caption{Combined VHE light curve from H.E.S.S.(triangles; $I(>300{\rm GeV})$; daily; average flux is 
$5.8\pm  1.7_{\rm stat}\pm 1.2_{\rm syst} \times 10^{-12}$ cm$^{-2}$ s$^{-1}$; \cite{Aharonian07b}) and MAGIC 
(diamonds;  $I(>150{\rm GeV})$; daily; average flux is 
$27.0\pm  1.7_{\rm stat} \times 10^{-12}$ cm$^{-2}$ s$^{-1}$; \cite{Albert07a}) of 1ES~1553+113. 
No indication of daily variability is apparent.
Dashed vertical lines correspond to the start and end times of the Suzaku observations. Inlay:
VHE light curve from H.E.S.S. during 24 and 25 July 2006 (run-by-run; average flux is 
$4.6\pm  0.6_{\rm stat}\pm 0.9_{\rm syst} \times 10^{-12}$ cm$^{-2}$ s$^{-1}$; \cite{Aharonian07b}).
No indication of variability is apparent on this time scale.
}
\label{fig6}
\end{figure}

\begin{figure}[t]
\resizebox{\hsize}{!}{\includegraphics{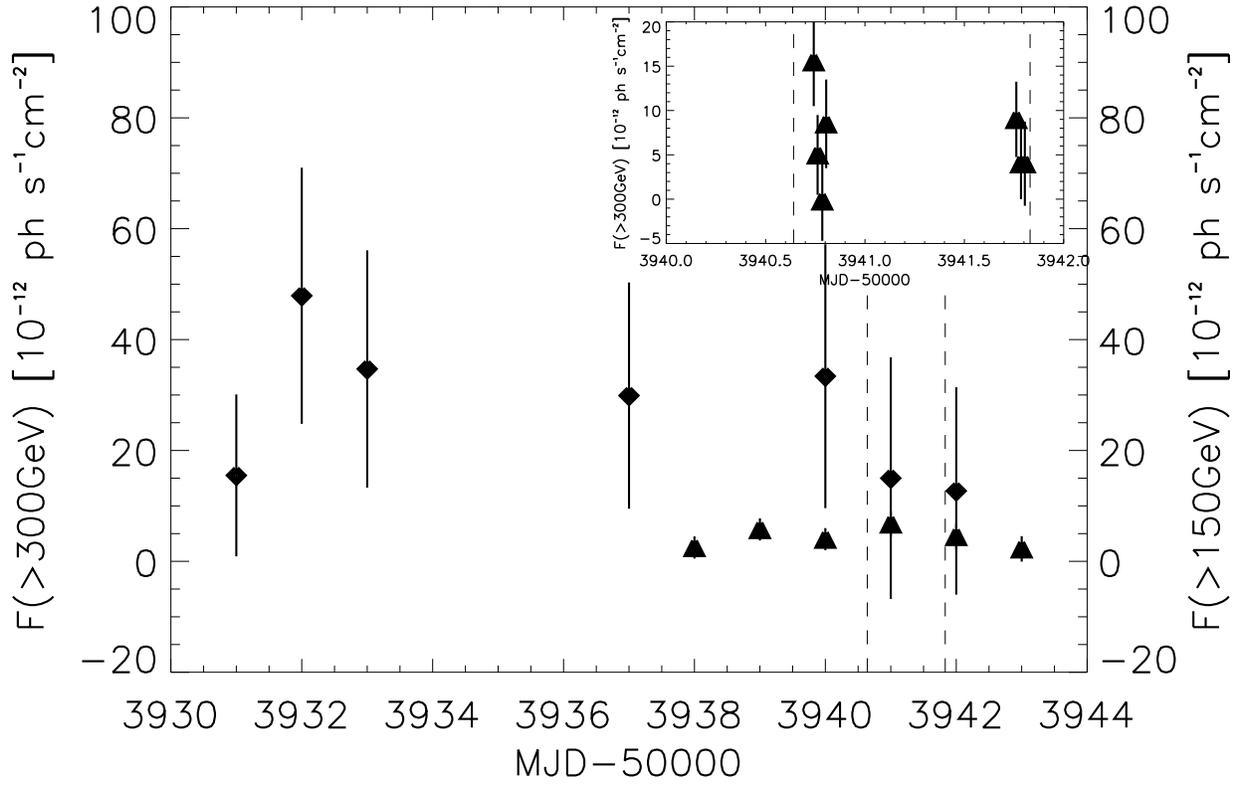}}
\caption{Observed VHE spectrum from 2005 (open symbols) H.E.S.S. (circles) and MAGIC (squares) observations of 1ES~1553+113, 
compared to the July 2006 VHE data \cite[filled symbols; ][]{Aharonian07b,Albert07a}. 
}
\label{fig7}
\end{figure}

The Suzaku observations of 1ES 1553+113 in July 2006 were also accompanied by simultaneous VHE observations, 
with H.E.S.S. and MAGIC. The total available live time data in July 2006 from H.E.S.S. encompasses 10.1 hours, out of which 3.1 hours live time 
observations were carried out simultaneously with the Suzaku pointing \cite{Aharonian07b}.
An excess of 101 events were found, corresponding to a significance of $3.9\sigma$. For the total July 2006 data
the significance increased to $6.7\sigma$ with an integrated flux of $I(>300{\rm GeV}) = (4.33\pm 0.94)\times 10^{-12}$ erg cm$^{-2}$ s$^{-1}$
\cite{Aharonian07b}.
A daily light curve did not show any convincing evidence of variability, as did a
run-by-run light curve from the Suzaku epoch observations (see Fig.~\ref{fig6}).
The low significance found by the H.E.S.S. observations during the Suzaku 
campaign did not justify to produce a reliable
fit to the $\gamma$-ray data \cite{Aharonian07b}. The whole July 2006 data 
set was reported with a power law fit with spectral photon index 
$\Gamma_\gamma=5.0\pm 0.7$. 
The spectral data points derived from this epoch are shown in Fig.~\ref{fig7}.
This is in very good agreement (within $2\sigma$) with the TeV-spectra 
derived from the past observations in 2005 and 2006.
The combined 2005/2006 data set is compatible with a spectral index of 
$\Gamma_\gamma \sim 4.9$ \cite{Aharonian07b}.

In July 2006 MAGIC was also observing 1ES~1553+113
for a total of 9.5 hours. These observations suffered from a 
sand-dust coming from the Sahara (Calima).  The effects of 
this weather condition on the data has been corrected 
using extinction measurements from the
Carlsberg Meridian Telescope \cite[][for details]{Albert07a}. 
With a live time of 8.5 hours the signal significance reached $5.0\sigma$
\cite{Albert07a}, enabling the compilation of a VHE spectrum for this time span.
The spectrum above $\sim 90$~GeV can be fitted 
by $dN/dE=(1.4\pm 0.3) 10^{-10} (E/200{\rm GeV})^{-4.1\pm 0.3}$TeV$^{-1}$s$^{-1}$cm$^{-2}$, 
and is in good agreement with the corresponding H.E.S.S. data points of July 2006.
For the 2 days of contemporaneous X-ray/VHE coverage the 
MAGIC data allowed only a $\sim 2\sigma$ result in 2.4~hours live time.
The spectral data points from these 2 days observations lie within the 
uncertainties of the overall July 2006 spectrum, 
which therefore will be used in the following as an adequate description 
of the VHE spectrum for the time of the Suzaku observation.

\begin{figure}[t]
\resizebox{\hsize}{!}{\includegraphics{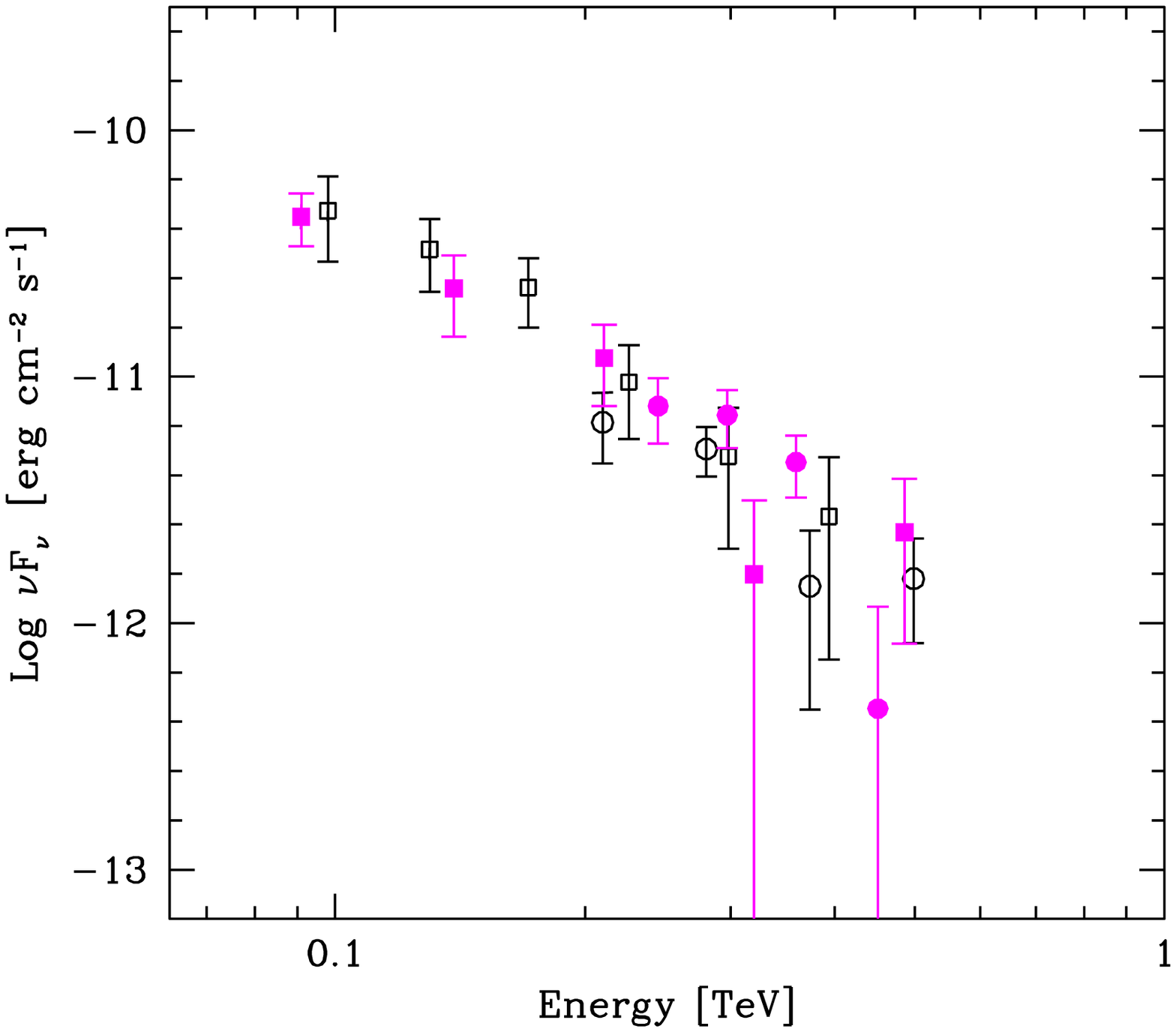}}
\caption{Combined VHE spectrum from H.E.S.S. \cite[black triangles;][]{Aharonian07b} and MAGIC  
\cite[black diamonds; ][]{Albert07a} of 1ES~1553+113. The combined fit (dashed line) follows a
power law $dN/dE = (2.192\pm 4.0742)\times 10^{-23} (E_{\rm TeV}/267.912)^{-4.068\pm 0.178}$cm$^{-2}$s$^{-1}$TeV$^{-1}$ ($\chi_{\rm red}^2=1.50$)
where $E_{\rm TeV}$ is the energy in TeV.
Green/light grey points belong to the de-absorbed VHE data using the minimal EBL \cite[P0.45 from][]{Aharonian06a}, 
red/dark grey points are for the \cite{Kneiske04} ``best-fit'' EBL-model and blue/medium grey points correspond to the \cite{Stecker06}
baseline EBL-model used for the de-absorption and assuming a source redshift of $z=0.3$.}
\label{fig8}
\end{figure}

Fig.~\ref{fig8} shows the spectral data points of the July 2006 MAGIC measurements together with the H.E.S.S. 
data points for this time span, covering the energy range of $\sim 90 - 500$~GeV.
A fit to the combined H.E.S.S. and MAGIC data points during the July 2006 observations revealed
a simple power-law representation to be adequate.
When compared to broken power law with exponential cutoff spectral shapes in the 90-500~GeV energy range, 
the goodness of fit does not improve noticeable.
The day-by-day light curve does not show any signs of variability (see Fig.~\ref{fig6}), 
although the average flux $> 150$~GeV
lies systematically above the corresponding H.E.S.S. $> 300$~GeV flux (because most of the VHE signal
stems from below $200-300$~GeV).
Also MAGIC confirms the surprising stability of the spectral shape of
this source, even when compared to year 2005 data \cite{Albert07a}.

Up to now 1ES 1553+113 is lacking any confirmed redshift measurement. The initially proposed redshift of $z=0.36$
turned out an mis-interpretation of line data \cite{Falomo90}. 
Constraints have been set only from indirect methods, e.g. 
from limits on host galaxy images \cite{Urry00,Treves07} one arrives at $z>0.25$. Recent spectroscopic observations improved
here to $z>0.09$ by deriving limits of the host galaxy absorption lines \cite{Sbarufatti06}. 
Gamma ray absorption in the EBL followed by pair production
in the VHE range offers a further possibility to
infer on redshift limits: The re-construction of the emitted $\gamma$-ray 
spectrum from the observed one with imposing limits 
on the hardness of the intrinsic source spectrum leads to the sought 
after redshift constraint. Such and similar concepts have been 
applied to past observations, and yielded $z<0.74$ \cite{Aharonian06b,Albert07a}, $z<0.69$ and $z<0.42$ \cite{Mazin07}.

In Fig.~\ref{fig8} we show the de-absorbed TeV-spectrum of 1ES 1553+113 for a source redshift of $z=0.3$ and
using various models for the EBL: the P0.45 flux level from Aharonian et al. (2006a), the
``best-fit'' model of Kneiske et al. (2004) and the latest version of the baseline model
of Stecker et al. (2006). The P0.45 flux level corresponds to 
a EBL flux density with a phenomenological spectral shape that describes the EBL data points satisfactorily on the level 
of the integrated galaxy counts.
All models support about the same $2\mu$m EBL-flux density, however, differ 
in their extension to lower and higher energies. Note that the baseline 
model flux at $<2\mu$m is not based on observational
constraints, but rather represents the extension of the Stecker et al. (2006) diffuse IR background calculation into
the optical/UV band, and should therefore be considered with caution in this energy range.
The resulting power spectra appear all at around photon index $\sim 3$ 
when de-absorbed by either model with the baseline model
providing the highest intrinsic VHE flux level up to $\sim 450$~GeV.
Still all considered EBL models lead to comparable photon powers delivered through the high and low energy 
SED humps. Stepping up in redshift the de-absorbed spectra are getting harder quickly.
At a source redshift of $z=0.67$ a spectral photon power law $\propto E^{-1}$ is reached at $>300$~GeV (see Fig.~\ref{fig8}), 
the asymptotically hardest spectrum from any conservative particle acceleration scenario.
We therefore consider a redshift of $z>0.7$ as unlikely for 1ES 1553+113 if the EBL flux density is
at the lowest possible value (integrated galaxy counts).
A search for the redshift that corresponds to an intrinsic $E^{-2}$ photon spectrum 
concludes at $z=0.5$ (see Fig.~\ref{fig8}). This represents the borderline for the VHE bump to lie at either tens of GeVs or $>$TeV.
When considering redshifts $<0.3$, the location of the VHE SED peak moves to $\leq 90$~GeV
with an estimated power output at $\gamma$-rays again comparable to the synchrotron power peak. 
The intrinsic photon spectral index is then $\sim 3.1$.

\clearpage

\begin{figure}[t]
\resizebox{\hsize}{!}{\includegraphics{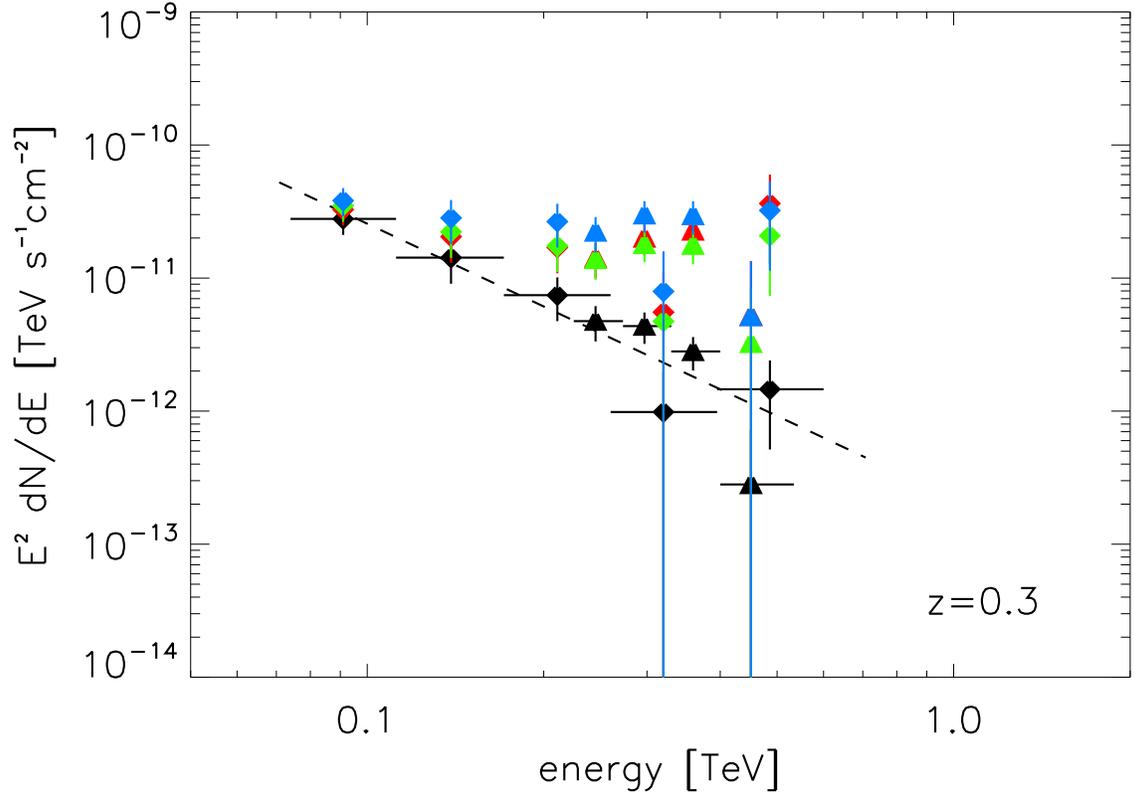}}
\caption{The intrinsic (de-absorbed) TeV-spectrum of 1ES 1553+113 for a source redshift of $z=0.67$ (blue/black),
$z=0.5$ (green/light grey) and $z=0.3$ (red/dark grey) using the minimal EBL \cite[P0.45 from][]{Aharonian06a}.
The source redshifts at $z>0.3$ have been chosen to lead to 
intrinsic $E^{-1}$ ($z=0.67$) and $E^{-2}$ ($z=0.5$) photon power law
spectra.}
\label{fig9}
\end{figure}
\clearpage

%%%%%%%%%%%%%%%%%%%%%%%%%%%%%%%%%%%%%%%%%%%%%%%%%%%%%%%%%%%%%%%%%%%%%%%%%%%%%%%%%%%%%%%%%%%%%%%%%%%%%%%%%%%%%%%%%%%%%%
\section{Discussion}

%%%%%%%%%%%%%%%%%%%%%%%%%%%%%%%%%%%%%%%%%%%%%%%%%%%%%%%%%%%%%%%%%%%%%%%%%%%%%%%%%%%%%%
\subsection{Historical X-ray behaviour of 1ES 1101-232 and 1ES 1553+113}

1ES 1101-232 has been detected during the past 15 years by a range of X-ray instruments, both covering the soft 
(ROSAT, XMM) and extending into the hard X-ray range (BeppoSAX, RXTE, SWIFT) at 
various sensitivity levels. 
Except for the continuous monitoring by RXTE's ASM and the 19.6~ksec monitoring by XMM \cite{Aharonian07a}, 
most observations were carried out in snapshot mode,
with each observation encompassing typically 1-5 ksec. The continuous Suzaku observation presented here 
constitutes therefore the first long ($>30$~ksec) high-sensitivity X-ray data set of this source.
None of the past observations showed significant short time variability ($<$hour). Although the continuous mode 
observations of Suzaku generally allow to probe intraday variability, we find no significant variability
of this time scale during our observations. Year-by-year variations 
as derived from the historical X-ray light curve are less than
a factor $\sim 2$. The highest ever measured X-ray flux level has been reported for 2005 from 
the RXTE-PCA measurements \cite{Aharonian07a}. There, in eleven consecutive nights the source 
was observed each night for $\sim 10$~ksec. 
No hints for strong flux variability were found, and the AGN was categorized 
to be in a non-flaring state. If this is the case, 1ES 1101-232 was likely 
never observed in a flaring state during the 
past 15 years. Only 4 months later SWIFT saw the source in a smooth decline of flux level with no 
sub-hour variations \cite[e.g.][]{Massa08}. Our Suzaku observations, 
made approximately $10$ months later, found the lowest ever measured X-ray flux 
from 1ES 1101-232.

So far, essentially all sensitive X-ray observations of 1ES 1101-232 prefer a curved or broken power-law 
representation instead a simple power-law \cite[e.g.,][]{Perlman05}, independent of flux level
and - as demonstrated in this work - extending into the hard X-ray range.
If interpreted as signatures of the particle energy gain process, one may conclude
that acceleration takes place at all flux levels observed so far.
In all cases there were no indications for a significant amount of X-ray absorbing material in the line-of-sight 
beyond the Galactic absorption column density.

The average spectrum during our Suzaku observations was quite similar to the one from
the 20\% higher flux XMM observations in 2001 \cite[$\Gamma\sim2.1-2.4$:][]{Perlman05}, 
where also no significant variability was noticed.  
The 1997/98 BeppoSAX observations \cite{Wolter00} 
had a similar value of the break energy at a factor $\sim 1.3-1.9$ higher flux level. 
The flux reported from the 2004 observations with XMM was comparable to the one measured by BeppoSAX
in 1997, however, the spectral break shifted to lower energies by a factor $\sim 1.2$.
Because of the lacking soft X-ray sensitivity, the PCA coverage of 1ES 1101-232 in 2005 revealed a break 
energy at much higher energies ($\sim 8$~keV). 
If the 1-8~keV X-ray spectral index is compared to the observed flux level at each observation, 
no statistically convincing indication for a flux-index correlation is found 
(Spearman rank correlation coefficient $R_s=-0.34$ with chance probability $P_c=0.33$, 
Kendall's $\tau_K=-0.30$ with $P_c=0.23$). Comparing, e.g., the XMM measurements carried out in 2001 and 2004, 
no significant spectral changes within the uncertainties occured despite a factor $\sim 1.6$ higher flux in 2004. 
On the other side, the BeppoSAX
observations revealed a spectral softening from the higher 
flux state in 1997 to the lower flux state in 1998. When compared to
the high flux RXTE data from 2005 a noticable softening occured despite the overall high flux level.
If following the evolution of the spectrum versus flux, no clear spectral hysteresis could be found.
In summary, it seems that spectral changes observed from annual visits of this source do not follow
particular patterns or relations with respect to flux changes. The overall spectral changes lie within
a rather narrow range between $\sim 2$ and 2.5 for the 1-8~keV photon spectral index.

The measured so far comparably long X-ray variability time scale in 1ES 1101-232 may imply that
this source was observed only in the quiescence over the last $\sim 15$~years. If one assumes 
the measured activity states
from one year to the other to be not causally related, the
annual measurements may then be treated as an independent ensemble of measurements from an HBL. In particular,
it seems reasonable to probe the behaviour in the peak luminosity-peak energy
diagram within the available HBL range. For a robust determination of the peak energy,
low energy data, preferably in the IR/optical/UV range, are necessary. Giommi et al. (2005) used
optical data from the literature in combination with the BeppoSAX measurements, while both XMM pointings had the 
advantage of simultaneous optical coverage with the Optical Monitor (OM) onboard the satellite, and SWIFT the 
onboard UVOT instrument. 
We did not find any convincing anti-correlation between peak energy and peak 
flux within the available 6 ensemble members
($R_s=-0.6$ with $P_c=0.21$, $\tau_K=-0.33$ with $P_c=0.35$).

1ES~1553+113 is known as a generally X-ray soft, bright HBL with only modest X-ray variability (factor $\leq 3.5$). 
SWIFT in 2005 \cite{Trama07,Massa08} achieved a detection of this source 
up to only 8~keV, despite its high flux level in October 2005 
that was reached within
half a  year from a 3.5 times lower flux level. BeppoSAX in 1998 
\cite{Donato05} saw it only with the LECS/MECS at a relatively low flux level, 
no detection with the PDS was possible.  With our Suzaku observation, 1ES~1553+113's spectrum
could finally be measured up to 30~keV while descending from the 2005 high flux level. 
Like for 1ES 1101-232, past X-ray observations encompass mostly snapshot 
observations of typically 3-10~ksec each. RXTE's observation 
campaign in 2003 \cite{Osterman06} involved visits to the source about 
3 times per day for $\sim 3$~ksec, for a total period of 21 days. 
During this time a very smooth rise in flux up to 3 times the
lowest flux level was observed. No hints of subhour variations were recorded in
any of the past observations. With our $\sim 50$~ksec continuous Suzaku measurements
we tested for intraday variability, but obtained a null result also at this intermediate flux level which 
was quite comparable to the 2001 flux recorded by XMM \cite{Perlman05}. 

No significant differences were found between the spectra reported from the 2001 and 2006
observations:
the average spectrum during our Suzaku observations was, within the uncertainties, in good agreement to the one from
the XMM observations in 2001 \cite[$\Gamma\sim2.2-2.4$: ][]{Perlman05}.
The surprising spectral stability of this source holds even for
the low flux state recorded with RXTE in 2003 \cite{Osterman06}, despite the factor $\sim$~5 difference
in flux. Even within the tripling of the flux during the PCA observations
spectral changes were not significantly beyond their uncertainties. In particular, any
systematic spectral trend with changing flux is not apparent, neither 
on a year-by-year time scale nor within the rising part of the RXTE monitoring data within their
admittedly large uncertainties: as is the case for 1ES 1101-232, the photon index above $3$~keV 
typically lingered around 2 and 2.5 even during flux increases by a factor $\sim 2$ or more. 
Those modest spectral changes together with the 
smoothness of the long-term light curve may suggest that explosive events are among the least
likely scenarios to account for the observed flux variations here.

The 1998 BeppoSAX observations \cite{Donato05} 
had a lower break energy ($\sim1$~keV) at a factor $\sim 2.5$ 
lower flux level as compared to the 2006 Suzaku observations.
SWIFT detected 1ES 1553+113 in both, a rather low flux state 
(April 2005) and relatively high flux level (October 2005), 
but with a surprisingly stable peak energy of $\sim 0.4-0.5$~keV in the XRT spectrum \cite[e.g.][]{Massa08}, 
$\sim 3$ times lower than the Suzaku data from 2006 imply.
The flux reached during the April 2005 observations with SWIFT was comparable to the one measured by BeppoSAX
in 1998, however the peak energy from the XRT data shifted to lower energies by a factor $\sim 2.5$.
No break energies were reported from the 2003 RXTE/PCA data \cite{Osterman06}.

Following the argumentation above we probe here the dependence between peak luminosity and peak energy
also for 1ES 1553+113, and utilize preferably data from the IR/optical/UV 
range to complement the X-ray band measurements.
Among the past observations only the XMM and SWIFT pointings were quasi-simultaneous measurements with optical/UV 
available, namely OM, UVOT and ROTSE data, respectively.
We use here our Suzaku observations with the data from the quasi-simultaneous 
optical coverage by the KVA optical telescope for a combined X-ray - optical spectral fit. 
A log-parabolic shape $dN/dE \propto E^{[-\Gamma-\beta\log(E)]}$
was used to determine the peak energy to $\sim 0.01$~keV with parameters $\beta=0.075\pm 0.001$ 
and $\Gamma=2.298\pm 0.003$. Note, however, that
this representation does only give a $\chi_{\rm red}^2 \approx 5.7$ due to the large 
weight given to the only optical point
(that has a small uncertainty) with respect to the many X-ray data points. 
More lower energy points would clarify this.
We did not use the peak values given for the 2003 RXTE campaign for this undertaking to
assure for a sample with equally determined peak energies:
the cm radio data from Osterman et al. (2006) for determining the peak energy would give too much weight
towards the longer wavelengths when compared to a procedure where only optical/UV 
in combination with the X-ray data
are used. The peak energies from the UVOT measurements range 
from $\sim 0.05$ to 0.02~keV from the high to the low flux state, 
respectively. The latter is in very good agreement with the values 
derived from the XMM 2001 observations.  Also for 1ES 1553+113 any convincing 
systematic correlation between peak energy and peak flux within the very limited sample of
4 ensemble members is lacking evidence ($R_s=0.39$ with $P_c=0.61$, $\tau_K=0.55$ with $P_c=0.26$).
This may indicate that particle acceleration up to the maximum energy is not preferentially limited by Compton losses 
if beaming does not change significantly.
Indeed, because of the lack of sufficiently dense radiation fields in HBL-type AGN, synchrotron
losses may dominate the radiative particle loss channel. The absence of a link between the 
peak luminosity and peak energy can then be understood if either an increase of the jet's synchrotron 
radiation field is not connected to an increase of the magnetic field strength alone on year time scales, or expansion 
losses and/or escape from the emission region sets a boundary to the maximum electron energy
at the so far observed activity states in 1ES~1101-232 and 1ES~1553+113.

\subsection{Spectral energy distributions}

\begin{figure}[t]
\resizebox{\hsize}{!}{\includegraphics{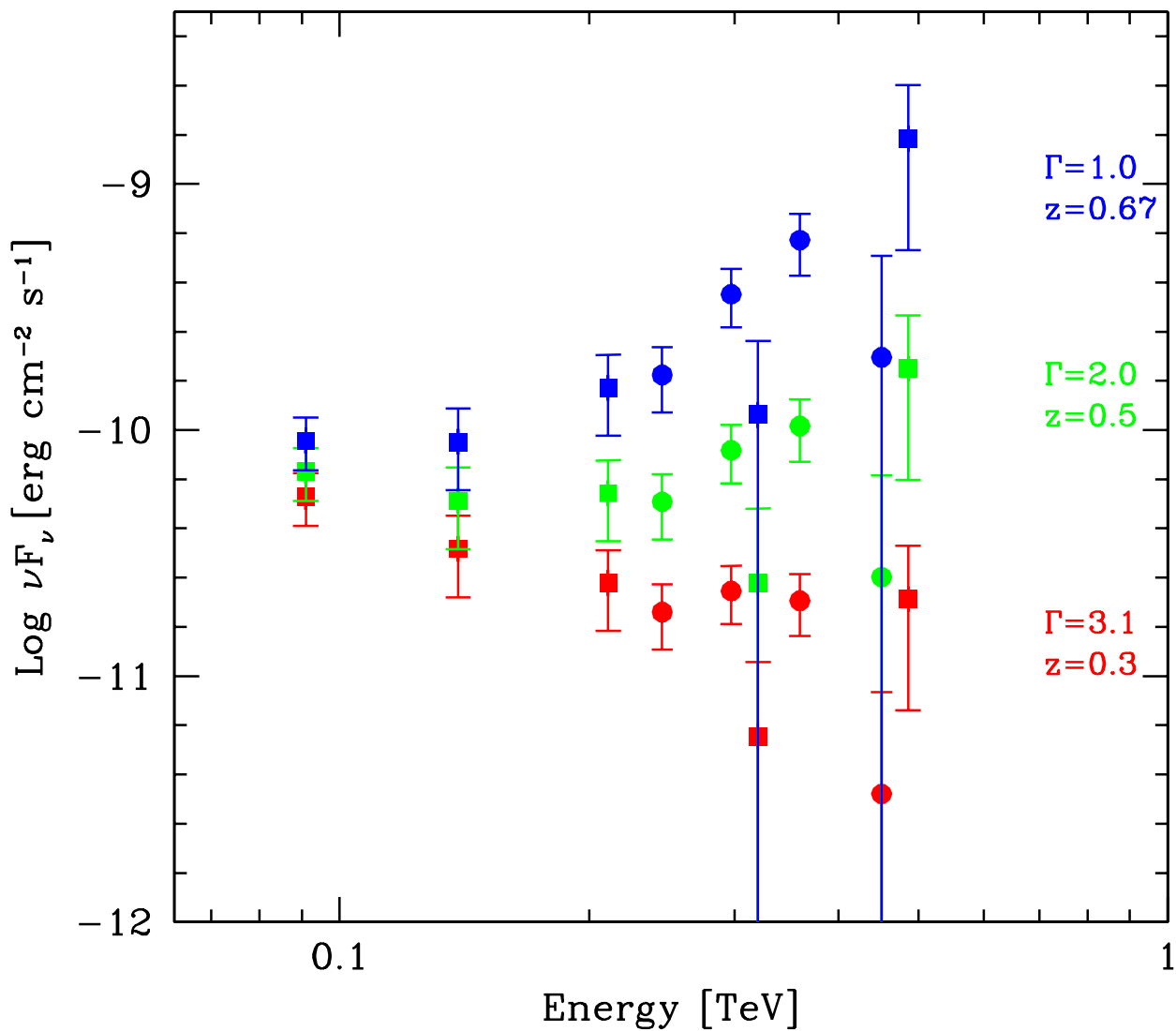}}
\caption{Broadband SED of 1ES~1101-232.
The observed H.E.S.S. data were de-absorbed using the minimal EBL \cite[P0.45 from][]{Aharonian06a}.
Present Suzaku data and the simultaneous de-absorbed 
July 2006 H.E.S.S. data are indicated in red. Blue points (XMM, RXTE, HESS) 
indicate data from 2004/05 \cite{Aharonian07a}, while historical
data collected from the literature are in black open symbols, and light grey \cite[BeppoSAX:][]{Wolter00}. The SWIFT data \cite[][not shown here]{Massa08} are 
compatible with the XMM flux level.}
\label{fig3}
\end{figure}

The overall SED of 1ES~1101-232, including the Suzaku
spectrum obtained by us and the simultaneous H.E.S.S. data is shown in Fig.~\ref{fig3}. 
If corrected by the minimal EBL \cite[P0.45 from][]{Aharonian06a}, comparable power is emitted
at $\gamma$-rays than at the synchrotron hump with a $\gamma$-ray peak beyond TeVs at the time
of the quasi-simultaneous Suzaku -- TeV observations. In the framework of a simple one-zone 
SSC model a (assumed uniform) field strength of sub-equipartition 
strength $B^2 \propto (\nu F_{\nu,IC})/(\nu F_{\nu,syn})$
can then be deduced. Alternatively, further target photon fields in the vicinity of the $\gamma$-ray emission
zone can lead to an enhanced $\gamma$-ray photon output. In hadronic blazar emission models TeV-emission
is the result of either reprocessed cascade emission initiated by pairs and/or photons produced in
photomeson interactions, or by proton and/or $\pi^{\pm}/\mu^{\pm}$-synchrotron radiation \cite[e.g.][]{Muecke01,Muecke03}.
In environments of weak ambient photon fields proton-photon interactions constitute a correspondingly
sub-dominant contribution to the $\gamma$-ray component, which then leaves proton synchrotron (or
curvature) radiation as the main $\gamma$-ray producer.

\clearpage
\clearpage

TeV observations of 1ES~1101-232 were used previously \cite{Aharonian06a} to set the most stringent limits
on the EBL at $\sim 0.7-4\mu$m using spectral considerations \cite{Aharonian06a}. These
limits could not be improved owing to the limited quality from the low flux TeV-data from 2006.

%%%%%%%%%%%%%%%%%%%%%%%%%%%%%%%%%%%%%%%%%%%%%%%%%%%%%%%%%%%%%%%%%%%%%%%%%%%%%%%%%%%%%%%%%%%%%%%%%%%%%%%%%%%%%%%%%%%%%%%%%%%%%%%%%%%%%%%%%%%%%

\begin{figure}[t]
\resizebox{\hsize}{!}{\includegraphics{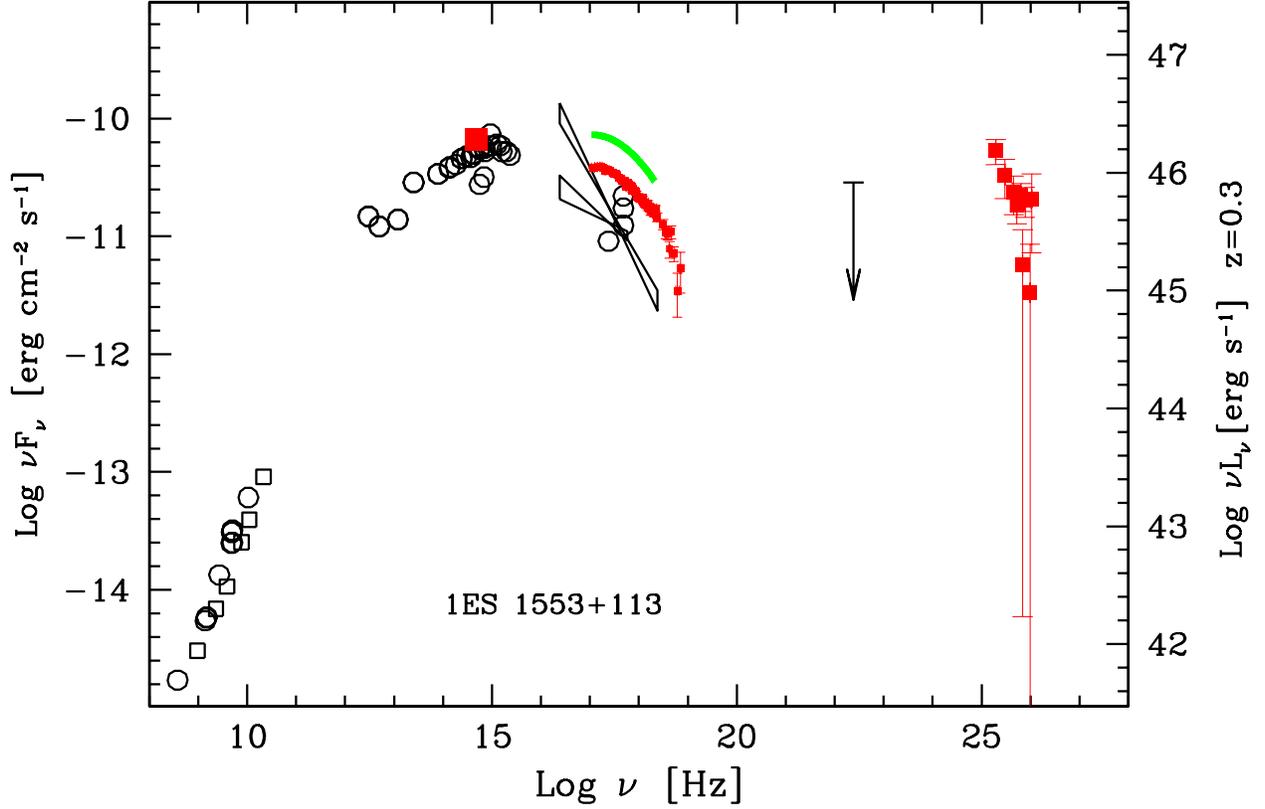}}
\caption{Broadband non-simultaneous SED of 1ES~1553+113.
The quasi-simultaneous data points (KVA, Suzaku, H.E.S.S., MAGIC) data are indicated as (red) squares. 
The green/grey line respresents the highest flux level observed by SWIFT \cite{Massa08} of this source.
The $\gamma$-ray data have been de-absorbed for a source redshift of $z=0.3$ 
using the minimal EBL \cite[P0.45 from][]{Aharonian06a}.}
\label{fig10}
\end{figure}

The overall SED of 1ES~1553+113 with the contemporaneous Suzaku, VHE (H.E.S.S., MAGIC) and optical (KVA) data is shown in Fig.~\ref{fig10}, complemented with historical ones from the literature.
Using a minimal EBL (P0.45) for de-absorption, approximately equal or more power is emitted
at $\gamma$-rays than at the synchrotron hump with a $\gamma$-ray peak at sub-TeVs.
In the framework of a simple one-zone SSC model, this ratio indicates sub-equipartition field strengths
in the emission region. Furthermore, if the same electrons emit TeV- and X-ray photons co-spatially, 
then the intrinsic spectral index at VHEs must not be harder than at X-ray energies.
For an EBL density at the galaxy counts flux level this is the case for $z \geq 0.4$.
Alternatively, external photon fields may contribute to
the target photon fields for inverse Compton scattering, or hadronic emission models may be
at work here. 

\clearpage
\clearpage

%%%%%%%%%%%%%%%%%%%%%%%%%%%%%%%%%%%%%%%%%%%%%%%%%%%%%%%%%%%%%%%%%%%%%%%%%%%%%%%%%%%%%%%%%%%%%%%%%%%%%%%%%%%%%%%
%%%%%%%%%%%%%%%%%%%%%%%%%%%%%%%%%%%%%%%%%%%%%%%%%%%%%%%%%%%%%%%%%%%%%%%%%%%%%%%%%%%%%%%%%%%%%%%%%%%%%%%%%%%%%%%
\section{Summary}

Continuous $\sim 50$ ksec Suzaku pointings towards each of the 
two distant TeV-blazars, 1ES 1101-232 and 1ES 1553+113, were 
carried out in May and July 2006, respectively, with quasi-simultaneous 
coverage at VHEs (H.E.S.S., MAGIC). In this work we presented the Suzaku data set, and set them into
context to the VHE observations and to their behaviour in the past. Our findings can be summarized as follows:

\begin{itemize}

\item Suzaku observations in 2006 of 1ES~1101-232 and 1ES~1553+113 found both
TeV-blazars in a non-active state. We detected both blazars from soft to hard X-rays up to $\sim 30$~keV, 
with 1ES 1553+113 measured to so far unprecedented energy in the synchrotron emission component.

\item The observed X-ray flux of 1ES~1101-232 is the lowest ever measured from this source.

\item The combined XIS and PIN spectra of both sources showed surprisingly similar spectral shapes.
This may indicate a common physical mechanism as being responsible for such spectral signature.
Both show indications of curvature, and can be equally well fitted by a broken power law
or log-parabolic shape. This curvature is found to extend well into the hard X-ray band ($\leq 30$~keV).
Note, however, that the PIN data of 1ES 1553+113 indicates a steepening beyond the log-parabolic or
broken power law model.

\item Simultaneous H.E.S.S. and MAGIC observations during the Suzaku epoch 
revealed no variability on any time scale probed by our data.  
H.E.S.S. obtained a weak detection ($\sim 3 \sigma$) of 1ES 1101-232 at the low flux level observed also during previous years.
The combined 1ES 1553+113 VHE spectrum ($90-500$~GeV) did not show any significant changes with respect
to historical VHE observations. Consequently, this data set did not allow to put further constraints on
the EBL flux density.

\item 1ES~1553+113 is unlikely to reside at a redshift $z>0.7$ even if the EBL is at the minimum flux density
(inferred from integral galaxy counts).

\end{itemize}

1ES 1553+113 and 1ES 1101-232, 
$\gamma$-ray blazars with surprisingly mild variability at virtually all 
energies, but nonetheless produce photons up to the highest TeV energies. 
This can be contrasted to the prominent TeV-blazars like Mkn~501, Mkn~421 
or PKS~2155-304 where rapid variations on subhour 
time scales at X- and TeV-energies have been detected at preferentially 
high activity states. Are there consequently various modes of
high energy production in blazar jets realized in nature -- rapidly explosive 
and quietly smooth ones -- , or are we being mislead by observations at
the instrument 
sensitivity limit and/or sparse sampling modes? Future high-sensitivity, 
dense sampling observations may answer this question.

%%%%%%%%%%%%%%%%%%%%%%%%%%%%%%%%%%%%%%%%%%%%%%%%%%%%%%%%%%%%%%%%%%%%%%%%%%%%%%%%%%%%%%%%%%%%%%%%%%%%%%%%%%%%%%%%%%%%%%%%%%%

\acknowledgments
AR acknowledges financial support from NASA grant NNX07AB19G, and GM by the 
Department of Energy contract to SLAC no. DE-AC3-76SF00515.
We thank the H.E.S.S.-, MAGIC- and KVA collaboration for their very constructive
support of this multifrequency campaign. We also thank the referee for a detailled report.

%%%%%%%%%%%%%%%%%%%%%%%%%%%%%%%%%%%%%%%%%%%%%%%%%%%%%%%%%%%%%%%%%%%%%%%%%%%%%%%%%%%%%%%%%%%%%%%%%%%%%%%%%%%%%%%%%%%%%%%%%%

%%%%%%%%%%%%%%%%%%%%%%%%%%%%%%%%%%%%%%%%%%%%%%%%%%%%%%%%%%%%%%%%%%%%%%%%%%%%%%%%%%%%%%%%%%%%%%%%%

\end{document}